\documentclass[doublecol]{epl2} 

\usepackage{amsmath}
\usepackage{graphicx}
\usepackage{dcolumn}
\usepackage{bm}
\usepackage{color}

\newcommand{\E}{\ensuremath{{\mathrm{e}}}}
\newcommand{\imag}{\ensuremath{{\mathrm{i}}}}

\title{Electron dynamics in graphene with gate-defined quantum dots} 

\author{A. Pieper, R. L. Heinisch and H.\ Fehske}
\shortauthor{A. Pieper, R. L. Heinisch and H.\ Fehske}

\institute{Institut f\"ur Physik,
             Ernst-Moritz-Arndt-Universit\"at Greifswald,
             D-17489 Greifswald, Germany}
             

\abstract{We use numerically exact Chebyshev expansion and kernel polynomial methods to study transport through circular graphene quantum dots in the framework of a tight-binding honeycomb lattice model. Our focus lies on the regime where individual modes of the electrostatically defined dot dominate the charge carrier dynamics. In particular, we discuss the scattering of an injected Dirac electron wave packet for a single quantum dot, electron confinement in the dot, the optical excitation of dot-bound modes, and the propagation of an electronic excitation along a linear array of dots. \vspace*{1pt}}

\begin{document} 

\maketitle
\section{Introduction}
Parallels between optics and electronics are particularly strong in graphene where low energy electrons have a linear energy dispersion. As a consequence, several unconventional electronic devices---modelled on optical analogues---have been proposed to control electron propagation in graphene. Most notably n-p junctions---described in the framework of geometric optics by a negative refractive index---could serve as Veselago lenses refocussing diverging electron rays \cite{CFA07}. Moreover, they show angle-sensitive transmission including---for normal incidence---perfect transmission \cite{CF06}. This so-called Klein tunnelling in graphene has been verified experimentally in a set-up of two n-p junctions forming a Fabry-Perot interferometer \cite{YK09}. For circular n-p junctions refraction inside the gated region leads to two caustics which coalesce in a cusp and focusses the electron density in the dot \cite{CPP07}. Both for planar and circular n-p junctions Rashba spin-orbit coupling gives rise to birefringence offering additional control on electron propagation \cite{BM10,AU13}.

While large gate-defined quantum dots are adequately described by ray optics, wave optical features emerge for smaller dots. In this regime resonances in the conductance \cite{BTB09,PAS11} and the scattering cross section \cite{HBF13a} indicate quasi-bound states at the dot. Indeed, electrons can be confined in a circular dot surrounded by unbiased graphene as the classical electron dynamics in the dot is integrable and the corresponding Dirac equation is separable  \cite{HA08,BTB09}. Arrangements of circular dots have also been studied as corrals forming a resonator \cite{VAW11} or in super-lattices where they renormalise the group velocity \cite{PYS08}. For a linear chain of dots, hybridisation of bound states gives rise to an effective hopping between dot-bound states \cite{HA09}.
  Most of these studies were performed using the continuum Dirac approximation and therefore have a limited range of validity. For planar junctions \cite{LBR12} and large dots \cite{PAS11} a tight-binding approach confirms results obtained in the Dirac approximation. For small dots, however, such a comparison remains to be done.

In this letter, we consider a lattice model and use exact numerical techniques to study the electron dynamics at circular gate-defined quantum dots. Thereby we confirm the scattering resonances and bound states for small dots. Specifically, we (i)  trace the time evolution of a wave packet scattered at a single dot, (ii) investigate the dot-bound normal modes, (iii) study the optical excitation of bound states at the dot, and (iv) follow the propagation of an excitation along a linear  chain of dots. 

\section{Model and Computational Method}
\begin{figure}[t]
  \centering
  \includegraphics[width=0.8\linewidth,clip]{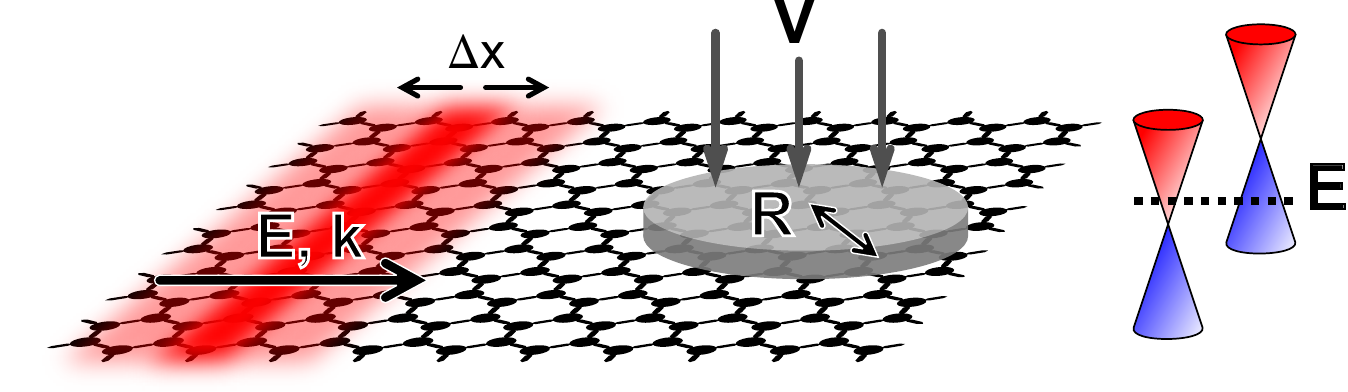}
  \caption{(Colour online) Schematic representation of the scattering geometry. An electron wave packet of  energy $E$, momentum ${\bf k}$ and width $\Delta x$ propagates along the zigzag direction in a broad graphene sheet. It impinges on a circular, electrostatically defined dot with radius $R$ and applied potential $V$. If the energy $E$ is close to the charge neutrality point, the Dirac approximation applies, yielding an linear dispersion. Note that the kinetic energy of the incoming 
  electron is positive (belonging to the conduction band), while it is negative inside the barrier  if $V>E$. Therefore the electron inside the quantum dot  belongs to the valence band, and a negative electron refraction index,
  determined by the wave vectors inside and outside of the gated region, is realised, $n=(E-V)/E$.}
  \label{fig:GFT_demo}
\end{figure}

Our investigations are based on the  tight-binding Hamiltonian
  \begin{equation}\label{H_bm}
     {H} =  \sum_{i}V_i^{} {c}_i^{\dag} {c}_i^{} 
           -t \sum_{\langle ij \rangle}({c}_i^{\dag} {c}_j^{} + c_j^\dag c_i^{})\,,
  \end{equation}
where $c_i^{(\dag)}$ is a fermionic annihilation (creation) operator acting on lattice site $i$ of a honeycomb lattice (with carbon-carbon distance $a=1.42$\AA), $\langle ij\rangle$ denotes pairs of nearest 
neighbours, and the site-dependent on-site potentials $V_i$ take values as appropriate for the system under consideration. The hopping matrix element $t\approx 3$ eV.  Actual gate-defined quantum dot devices might be realised by external electrostatic fields (gates, barriers). Basically we assume the potential $V_i=V$ for sites inside the dot and zero outside this region (see Fig.~\ref{fig:GFT_demo}).  However, in order to guarantee a smooth potential on 
the scale of the lattice constant, we adopt a linear interpolation of $V_i$ within a small range $R\pm 0.01R$ (except stated otherwise). 

The time propagation and scattering of a Gaussian wave packet created at time $\tau_0$ with width $\Delta x$ and momentum $k_x$  (cf. Fig.~\ref{fig:GFT_demo}), can be obtained by expanding---in $|\psi(\tau)\rangle= U(\tau,\tau_0) |\psi(\tau_0)\rangle$---the time propagation operator $U(\tau,\tau_0)=U(\Delta \tau)$   into a finite ($M$) series of first-kind Chebyshev polynomials  $T_l(x)=\cos(l\, \mathrm{arccos}(x))$ \cite{FSSWFB09}:
$  U(\Delta \tau) =  \E^{-\frac{\imag b \Delta \tau}{\hbar}}   \left[ c_0\left(\frac{d\Delta \tau}{\hbar}\right) + 2\sum\limits_{l=1}^{M} c_l\left(\frac{d\Delta \tau}{\hbar}\right)    T_l(\tilde{H}) \right]\,,$
where  $c_l (d \Delta \tau/\hbar)  =  (-\imag)^l J_l (d \Delta \tau/\hbar)$. $J_l$ is the Bessel function. The spectrum of $\tilde{H}=(H-b)/d$ is rescaled by $b$ and $d$  to be within the interval $[-1,1]$.
This method has also been applied to wave-packet dynamics in graphene \cite{KT09,SF12}.

Then the time-dependent (local) particle densities and current densities can be obtained numerically from
\begin{eqnarray}
n_i(\tau)&=&\left|\langle i| U(\tau, \tau_0) |\psi(\tau_0)\rangle\right|^2\,,\\
\vec {j_i}(\tau)&=&|\langle i| (- \vec J\,/e) U(\tau, \tau_0) |\psi(\tau_0)\rangle|^2 
\end{eqnarray}
with the charge current operator  \begin{equation}
\vec J = -\frac{{\rm i}te}{\hbar}\sum_{\langle i j\rangle}(\vec {r_j} - \vec {r_i}) ( c_i^\dagger c_j^{}- c_j^\dagger c_i^{} )\,. \label{eq.currop}
\end{equation}

\section{Scattering by a single dot}

\begin{figure*}[t]
  \centering
  \includegraphics[width=0.84\linewidth,clip]{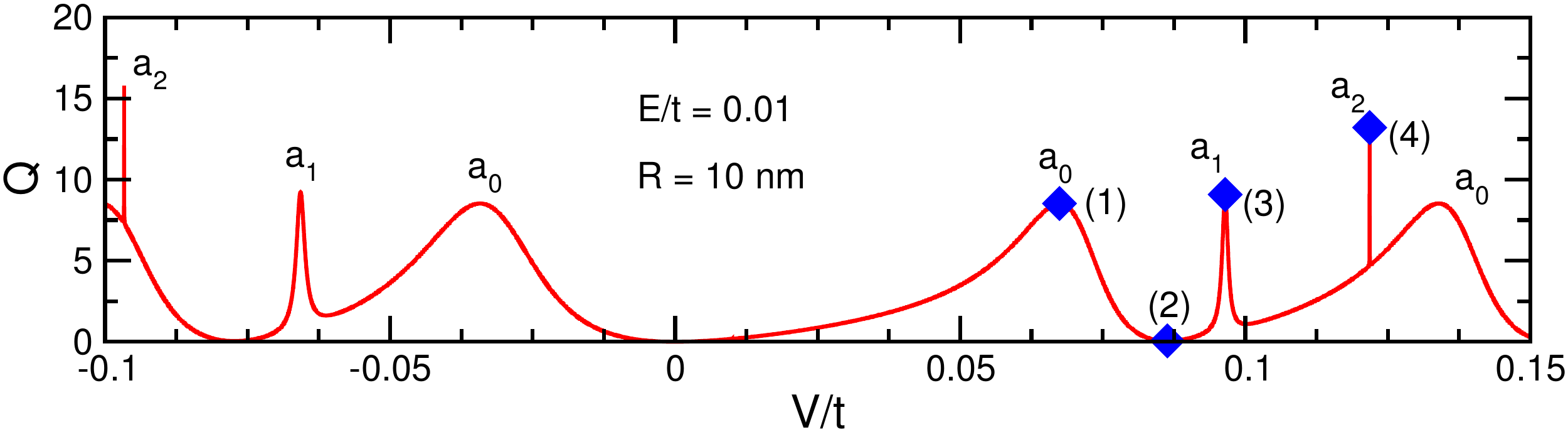}\\
\includegraphics[width=0.85\linewidth,clip]{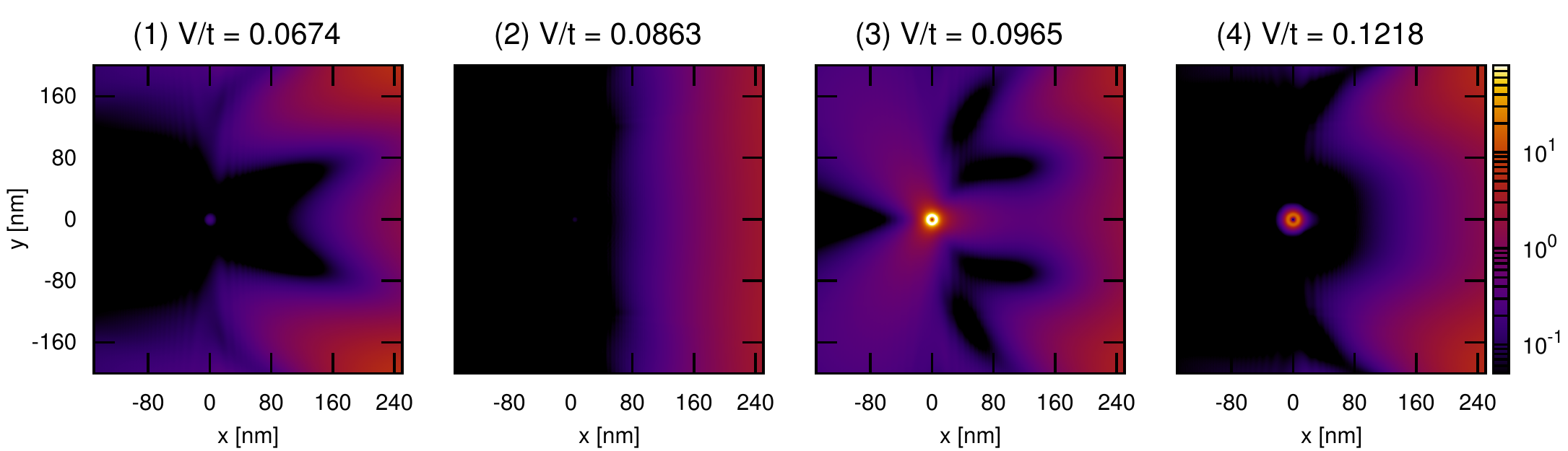}\\
 \includegraphics[width=0.85\linewidth,clip]{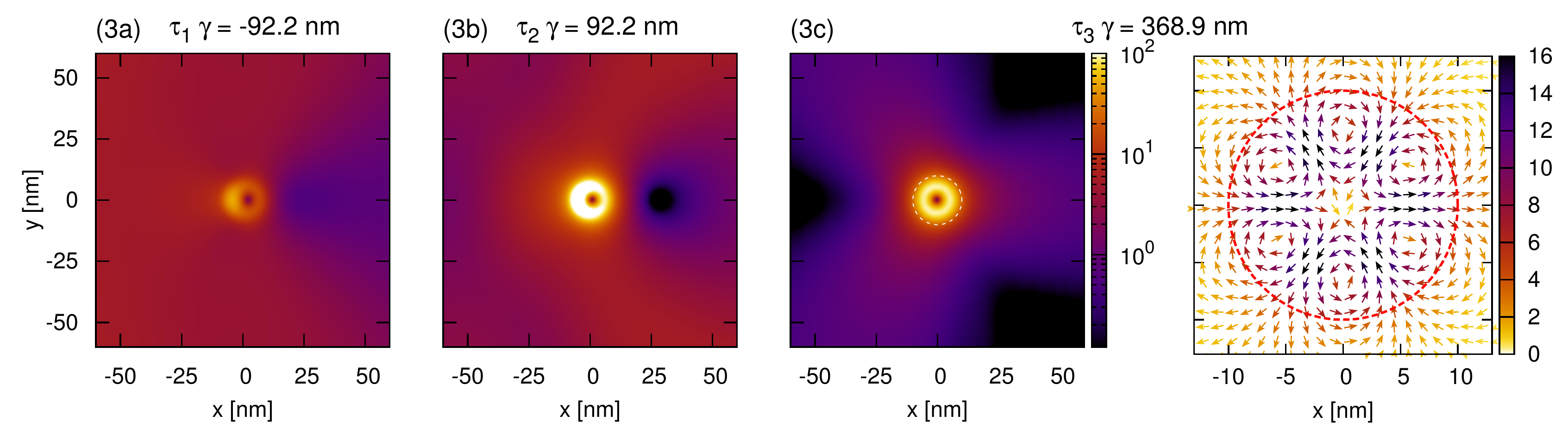}
\caption{(Colour online) Scattering and particle confinement by a gate-defined circular quantum dot. The upper panel shows the efficiency $Q$ for plane wave scattering in the Dirac approximation. Depending on the potential $V$ (at fixed energy $E/t=0.01$ and dot radius $R=10$ nm) scattering resonances appear when specific modes $a_m$ are excited~\cite{HBF13a}. Below, density snapshots obtained numerically for a tight-binding Hamiltonian are shown for wave packet scattering on a graphene sheet with $N_z\times N_a=12.000 \times 4.000$ sites ($z$ denotes the zigzag and $a$ the armchair direction) corresponding to $(1476 \times 852)$~nm$^2$. The wave packet with width $\Delta x = 148$~nm is initialised at the position  $\tau_0 \gamma=-369$~nm ($\gamma=1.5 a t /\hbar$ is the slope of the energy dispersion at the Dirac point). The middle panels give the electron density for specific $V$ [marker (1)-(4) in the upper panel] at  time $\tau_3=369$~nm$/\gamma$ when the wave packet has passed the dot already.  The lower panels show three time steps during the scattering process at the $a_1$ resonance [marker (3)]. Here we also show the current density. The dot is indicated by the dashed circle.} 
   \label{fig:mie_scattering_new}
\end{figure*}

We begin our analysis with the time-evolution of an electron scattered at a single quantum dot. Figure \ref{fig:mie_scattering_new} shows our results for $E/V<1$ (entailing negative refractive index). For comparison with the continuum model for plane wave scattering, we plot in the upper panel the scattering efficiency $Q$, that is the scattering cross section divided by the geometric cross section, calculated in the Dirac approximation (for details of the calculation see \cite{HBF13a}). A series of resonances $a_m$ appears in $Q$, due to the excitation of normal modes of the dot. The modes $a_0$ turn out to be rather broad, while higher modes $a_{m>0}$ entail very sharp resonances in $Q$. 
In the regime of resonance the mismatch of wave length---large outside  and small inside the dot---suppresses the excitation of normal modes except for the cases where one normal mode fits particularly well into the dot for a specific combination of $E$, $V$ and $R$. Hence, resonances also appear for $V<0$,  that is an n-n junction (entailing positive refractive index), where these criteria can be satisfied too. 

We compare this with the full lattice calculation for specific bias potentials $V$ at the dot [marked (1) to (4) in Fig.  \ref{fig:mie_scattering_new})]. For this, we consider the scattering of a wave packet at the dot. The middle panels show density snapshots taken a long time after the wave packet has passed through the dot. If scattering is dominated by the $a_0$ mode (marker 1) a significant fraction of electron density is removed from the incident wave. Indicated by the broad line shape of the $a_0$ mode in $Q$ (upper panel) electrons are only trapped for a short time at the dot. As the major part of density removed from the incident wave has been scattered off before the time of the density snapshot only a small increase of density at the dot can be seen. For low $V$ only the $a_0$ mode can be excited. If this is not the case (marker 2) the wave packet passes undisturbed through the dot. Tuning $V$ on the resonance of $a_1$ (marker 3), the density snapshot shows a significant dent in the passing wave front together with strong particle confinement at the dot. Increasing $V$ further resonant scattering by the $a_2$ mode overlaps with a broad $a_0$ background (marker 4). As a consequence, a substantial fraction of density is removed from the wave packet and scattered off (due to the mode $a_0$) and another substantial fraction is trapped at the dot (due to the mode $a_2$) from which it is scattered off very slowly.

In the lower panel of Fig. \ref{fig:mie_scattering_new} we exemplify the time evolution during a scattering event at the $a_1$ resonance (marker 3). Upon impact of the wave packet on the dot density is collected at the dot. This leaves a depleted wake behind the dot. The trapped fraction of density is ring-shaped but not yet fully symmetric. After the passage of the wave packet, this transient behaviour ends and the symmetric ring-shaped density profile, which perfectly fills the dot, fully develops. The current field shows six vortices in agreement with the Dirac approximation. They are responsible for electron trapping at the dot. The density profile shows three directions with increased electron density corresponding to the three preferred scattering directions of the $a_1$ modes (also seen in the current field). 

\begin{figure}[t]
  \centering
  \includegraphics[width=\linewidth,clip]{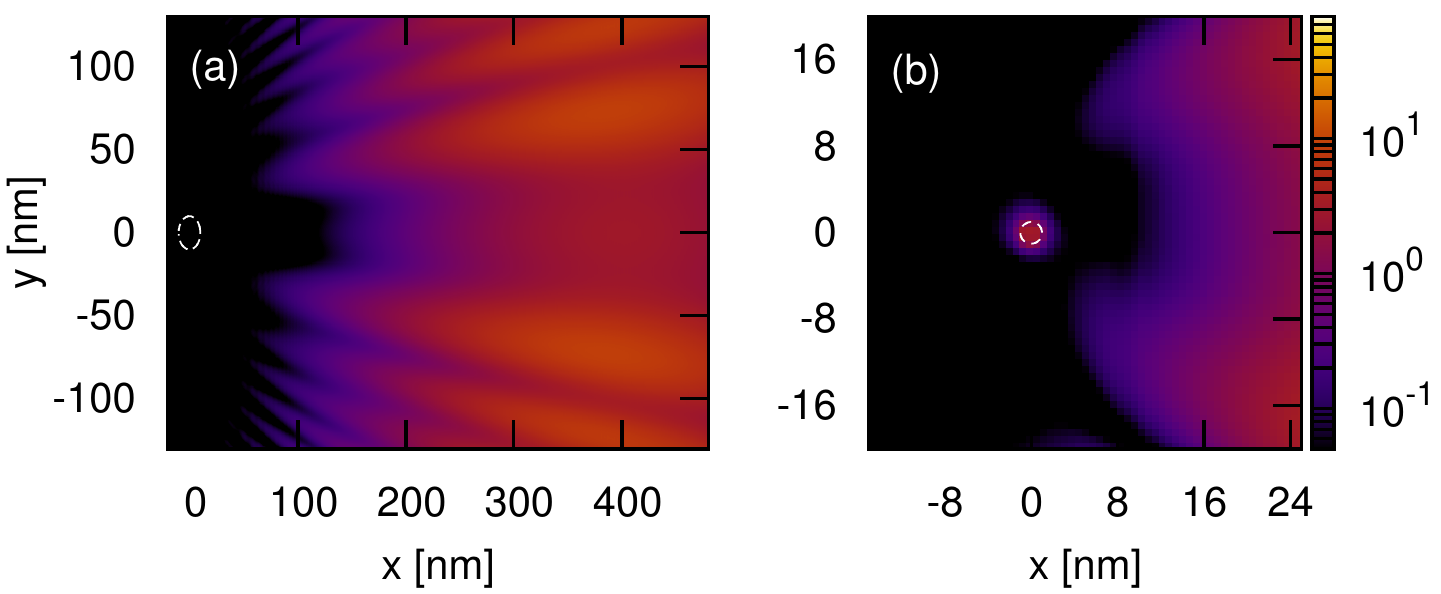}\\
 \caption{(Colour online) Density snapshots after wave packet scattering on a circular dot. Left: Threshold incidence  with $E/t=V/t=0.09648$ (Sample size, dot radius, initialisation of the wave packet and time of the snapshot are chosen as in Fig. \ref{fig:mie_scattering_new}). Right: Mode $a_1$ away from the Dirac regime for a very small dot with $R=1$ nm, $E/t=0.1$, $V/t=0.9648$, $\tau_0 \gamma=-36.9$ nm, $\tau \gamma=36.9$ nm and $\Delta x=14.8$ nm.} 
   \label{fig:overbarrierscatt}
\end{figure}

For larger $E$ an increasing number of modes $a_m$ contribute to scattering. An interesting particular case in this respect is $V=E$ (see Fig. \ref{fig:overbarrierscatt}a). In this case the energy of the incident electron falls on the Dirac point inside the dot. No temporary electron trapping takes place. Instead, diffraction at the dot leads to an electron depleted wake behind the dot and to an interference pattern in the passing wave packet. 

Provided both $E$ and $V \ll t$ the lattice calculation confirms results obtained within the Dirac approximation. If either  $E$ or $V$ is not small compared to $t$,  deviations from the linear energy dispersion become appreciable. In Fig.  \ref{fig:overbarrierscatt}b we consider the resonance $a_1$ for a very small dot which entails $V\approx t$. The resonance is---albeit weakened---still observable. The particle confinement, however, is much weaker and characteristics such as the ring-shaped density and the three characteristic scattering angles are no longer discernible.

\section{Properties of dot-bound normal modes}

\begin{figure}[t]
\begin{center}
  \includegraphics[width=0.65\linewidth,clip]{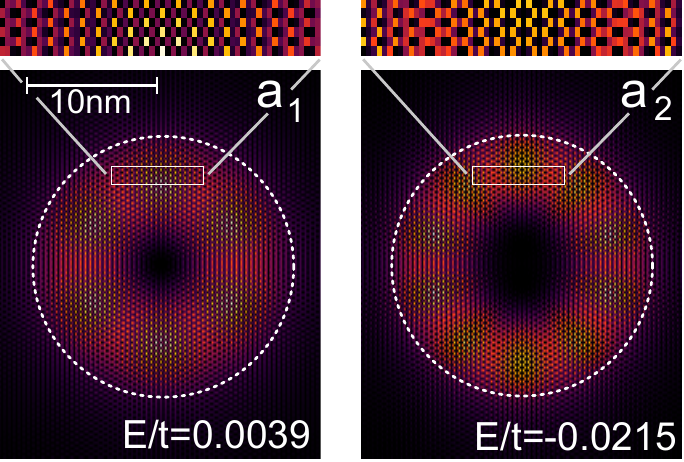}
\includegraphics[width=0.33\linewidth,clip]{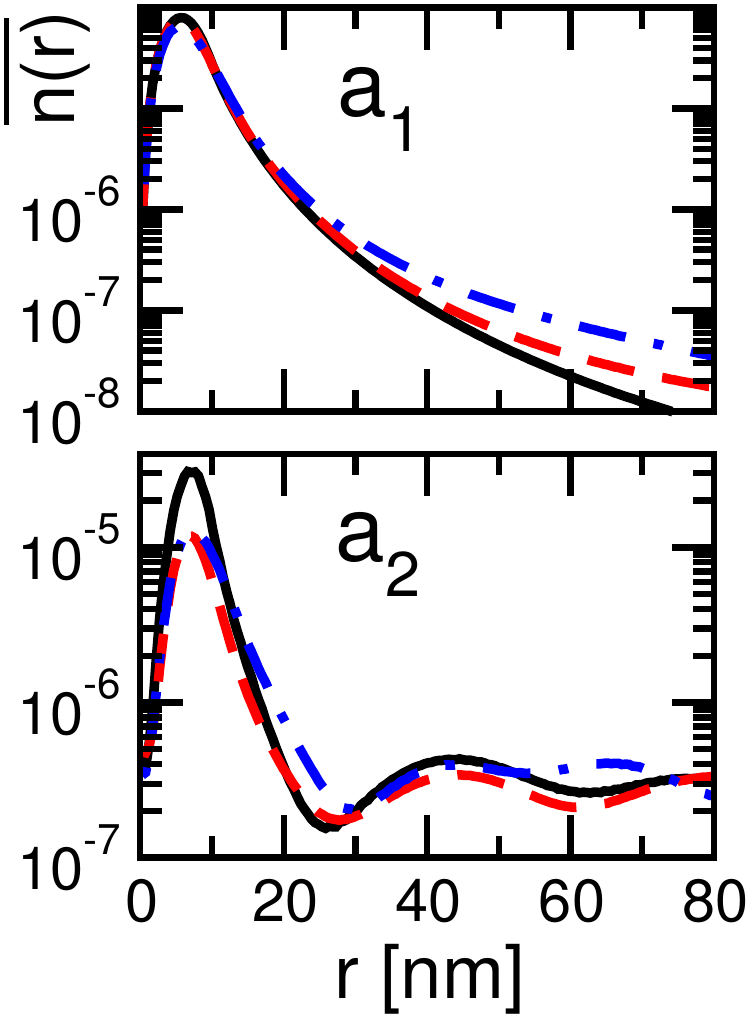}
\caption{(Colour online) Dot-localised states associated with the normal modes $a_m$. Left: density profiles obtained by ED for a graphene sheet with periodic boundary conditions and $N_z\times N_a=2000\times 1200$ corresponding to (246 $\times$ 256) nm$^2$ with a dot with $R=10$ nm and $V/t=0.081615$. Each panel shows one representative out of the four states making up a mode $a_m$. The intensity, normalised to the maximum density, increases from black over red to yellow. Magnifications are shown above. Each tile gives the electron density on one lattice site. Right: Radial density profile for different grading widths $R\pm0.01R$ (solid line), $R\pm0.3R$ (dashed line) and $R\pm0.5R$ (dash-dot line). The modes $a_m$ are only weakly sensitive to the grading width. The mode $a_1$ is a bound state, while the mode $a_2$ is a long-living decaying state.   }
\label{fig:ED_states}
\end{center}

\begin{center}
\includegraphics[width=0.8\linewidth,clip]{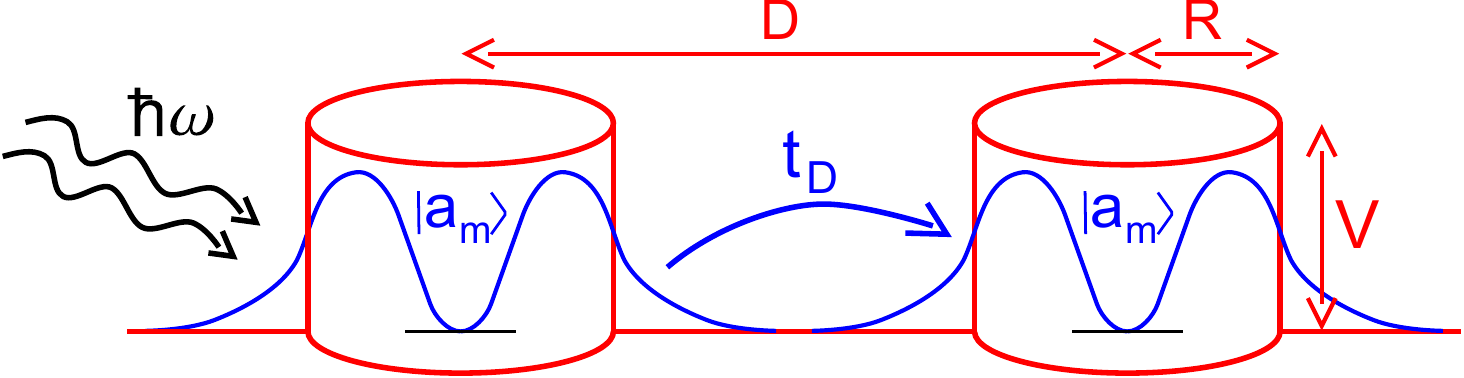}
\caption{(Colour online) Schematic representation of a linear array of dots. Dots with radius $R$ have a distance of $D$. The applied bias $V$ is such that a free standing dot supports a bound state with energy $E\approx0$. Overlap with bound states at neighbouring dots leads to a hybridisation of bound states and gives rise to an effective dot-dot transfer (indicated by $t_D$). The dot-localised states can also be probed optically.
}
\label{fig:dotstates}
\end{center}
\end{figure}

Normal modes $a_m$ of the dot lead to sharp resonances in the scattering efficiency. They appear for particular values of $R$, $V$, and $E$ and allow electron trapping at the dot. From the perspective of scattering the energy of normal modes can be determined, for given $V$ and $R$, by setting the imaginary part of the denominator of the scattering coefficients zero \cite{HBF13a}.

 For an equilibrium situation (without incident wave) the normal modes can be interpreted as decaying states. In this case they are the solutions of the Dirac equation with outgoing boundary condition and complex energies \cite{HA08,HA09}. The life-time of these quasi-bound states can be very long and for the case $E=0$ and particular combinations of $V$ and $R$ they become true bound states \cite{BTB09}. For $E\neq 0$ the dot-bound states hybridise with extended states and an electron can only be trapped temporarily by them. 

Each mode $a_m$ consists of two states. One has the total angular momentum (orbital angular momentum and pseudo-spin) of $j=m+1/2$, the other of $j=-m-1/2$. Over the course of wave packet scattering both are excited and their superposition  gives the characteristic vortex pattern of the modes $a_m$ (see Fig.~\ref{fig:mie_scattering_new} for $a_1$). As these two states are twofold degenerate with respect to the valleys $K$ and $K^\prime$, the mode $a_m$ is associated with four states.

We study the bound or quasi-bound states for a single dot on graphene by means of exact diagonalisation (ED). The energy spectrum we obtain consists of extended states, which show no enhanced density at the dot, interspersed with groups of four states having very small energy differences and large density at the dot. They can be identified  as the normal modes of the dot. One representative of the four states making up the modes $a_1$ and $a_2$  is shown in Fig.~\ref{fig:ED_states}. Here, we have chosen $V$ and $R$ such that the mode $a_1$ at $E\approx 0$ is a bound state and the mode $a_2$ is a quasi-bound state with $E<0$. 

The density profiles of the states obtained by ED show a pattern in the sub-lattice make-up which can be related to the vortex structure in the current field at the scattering resonances. Comparing for the mode $a_1$ the density profile from ED (Fig.~\ref{fig:ED_states}) with the current field  (Fig.~\ref{fig:mie_scattering_new}) we find that at the vortex cores only one sub-lattice is occupied while between them the electron density is distributed equally on both sub-lattices. Averaged over several lattice sites the density is radially symmetric. 
A superposition of the state shown in Fig.~\ref{fig:ED_states} with another state of the $a_1$ mode which also occupies the same sub-lattice at the vortex cores realises the current pattern shown in Fig.~\ref{fig:mie_scattering_new}. In the vortex cores only one sub-lattice is occupied. Thus, there can be no hopping between neighbouring lattice sites and the current is zero. Between the vortex cores both sub-lattices are occupied and hopping  gives rise to a current. Figure~\ref{fig:ED_states} also confirms that the character of the modes $a_m$ is relatively insensitive to the grading width of the potential.

\begin{figure*}[htbp]
\centering
{\includegraphics[width=0.9\linewidth,clip]{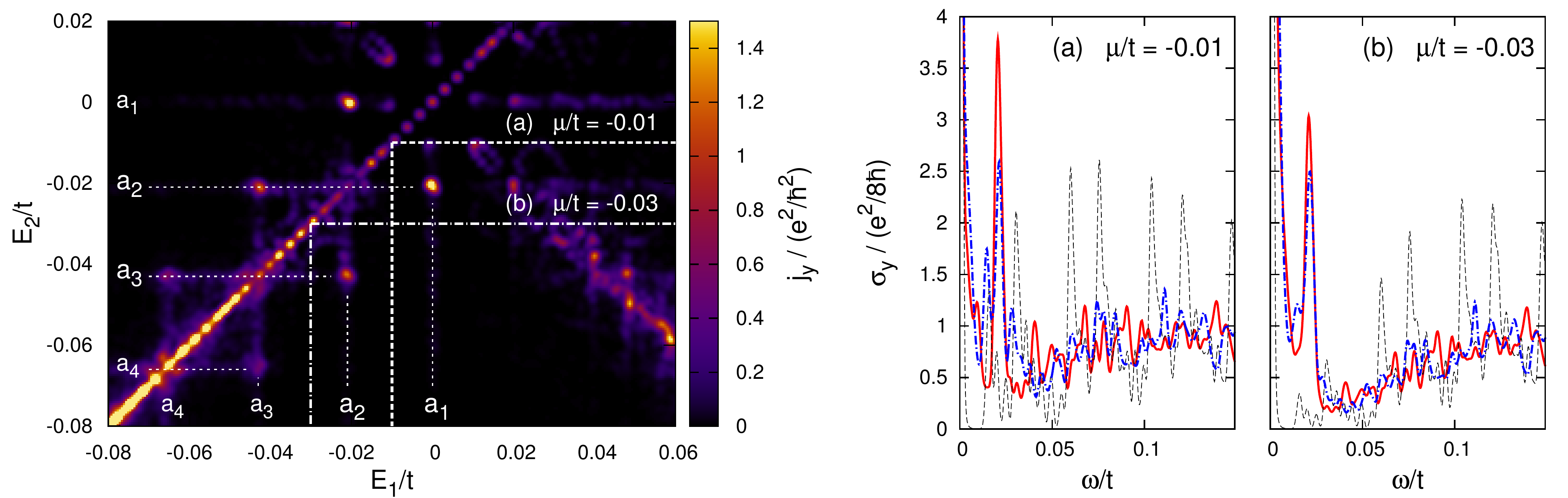}} \\
 \caption{(Colour online) Left: Current matrix element $j_y(E_1,E_2) $ as a function of the energies $E_1$ and $E_2$ for a chain of ten dots in an armchair graphene nano-ribbon with $N_z\times N_a=488\times1888$ corresponding to (60 $\times$ 402) nm$^2$. The dot radii $R=10$ nm, their separation $D=40$ nm, and the potential $V/t=0.081615$. Right: Optical conductivity of this system (full line), a typical sample of 10 dots with a random distribution of radii $R \in [9.5$nm,$10.5$nm$]$ (dash-dot line) and a nano-ribbon without dots (thin dashed line) for two different chemical potentials $\mu$ at $\beta=100/t$ (corresponding to $T=350$ K). The Drude peak at $\omega=0$ is due to $\mu<0$. The domain of the current matrix element which contributes most is indicated by the dashed lines. In (a) the peak at  $\omega/t=0.02$ is due to transitions from the $a_2$ to the $a_1$ mode. In (b) this peak corresponds to the $a_3$ to $a_2$ transition. Varying the chemical potential---by tuning the voltage at the back gate---one can bring different transitions $a_{m}\rightarrow a_{m\pm1}$ into observational range. The signature of these transitions does not change much even for slightly different dot radii.}   
   \label{fig:opt}
\end{figure*}

\section{Optical excitation of  dot-bound normal modes}

In the previous sections we have studied the excitation of quasi-bound states at the dot during wave packet scattering. For the case $E=0$, where the dot-bound states do not decay, they are not coupled to the continuum and can not be populated in the course of a scattering process. These dot-bound states could be excited optically. Signatures of such a process should be found in the optical conductivity which we will study in the following. 
The so-called regular contribution to the real part of the optical conductivity is given by
\begin{align}
\sigma_\alpha(\omega) =&\frac{\pi \hbar}{\omega \Omega} \sum_{i,j}  | \langle i| J_\alpha | j \rangle |^2 [f(E_i)-f(E_j)]  \delta(\omega +E_i-E_j) \nonumber \\ 
=& \frac{\pi \hbar}{\omega} \int_{-\infty}^\infty \mathrm{d}E j_\alpha(E,E+\omega) [f(E)-f(E+\omega)] 
\end{align}
with $\Omega=N_a N_z \sqrt{3} a^2/4$. $f(E)=\left[e^{\beta(E-\mu)}+1 \right]^{-1}$ is the Fermi function with chemical potential $\mu$. The current operator $J_{\alpha=x,y}$ is given by Eq.~(\ref{eq.currop}). $\sigma_\alpha(\omega)$ can be efficiently calculated by the kernel polynomial method \cite{WWAF06,PSWF13}. The key information is encoded in the current matrix element 
\begin{equation}
 j_\alpha(E_1,E_2)=\frac{1}{\Omega} \mathrm{Tr} [J_\alpha \delta(E_1-H) J_\alpha \delta(E_2-H)]\,.
\end{equation}

As the effect of one dot in a large graphene sheet is relatively small we study the optical conductivity for an array of dots. Specifically, we consider a linear chain of ten dots on an armchair nano-ribbon. Again we adjust $V$ and $R$ so that a free-standing dot supports a bound $a_1$ mode. For the linear chain of dots (illustrated in Fig.~\ref{fig:dotstates}) the coupling of the normal modes at neighbouring dots broadens the bound states somewhat.

The current matrix element is shown on the left panel of Fig.~\ref{fig:opt}. The spectral weight along the diagonal $E_1=E_2$ controls the DC conductivity ($\omega=0$). For an armchair ribbon without dots (not shown) no spectral weight appears close to $E_1=E_2=0$ as a consequence of a mid-band energy gap caused by finite-size effects. For a system with dots, however, some spectral weight is found in this region. Spectral weight along the line $E_1=-E_2$ stems from the vertical $\pi$--$\pi^\ast$ inter-band transitions and contributes to the AC conductivity. 

Transitions between dot-bound states lead to sharp, selective increases in $j_y(E_1,E_2)$ if both $E_1$ and $E_2$ coincide with energies of normal modes $a_m$. Calculating the current matrix element between two normal modes $|m_1\rangle$ and $|m_2\rangle$ in the Dirac approximation we find $\langle m_1|J_\alpha |m_2 \rangle \sim \delta_{m_1,m_2 \pm 1}$. This selection rule, which allows only transitions $a_m \rightarrow a_{m \pm 1}$, is confirmed by Fig.~\ref{fig:opt}. 

While the current matrix element is not directly accessible experimentally, the optical conductivity is (shown on the right panel of Fig.~\ref{fig:opt}). For a narrow nano-ribbon without dots, van Hove singularities, linked to the set of quasi one-dimensional bands, lead to sharp peaks in $\sigma_y(\omega)$. For a ribbon with dots these peaks are suppressed. At low energies, the optical conductivity shows a sharp peak at the energy difference between two dot-bound modes which reflects transitions $a_m \rightarrow a_{m\pm 1}$. The frequency of these transitions is $\omega \propto \gamma /R$ (in the Dirac approximation we obtain $\omega_{n,m} \simeq \hbar \gamma (j_{m,1}-j_{n,1})/R$ with $j_{m,s}$ zeros of Bessel functions). Tuning the chemical potential $\mu$, different transitions between dot-bound states can be selected. Note that the signature of transitions between dot-bound states prevails even for a uniform distribution of radii by 10\%. Thus an absorption measurement for an array of dots on a nano-ribbon should allow to probe dot-bound states optically.

\section{Transport through a linear array of dots}

\begin{figure*}[t]
\centering
\includegraphics[width=0.9\linewidth,clip]{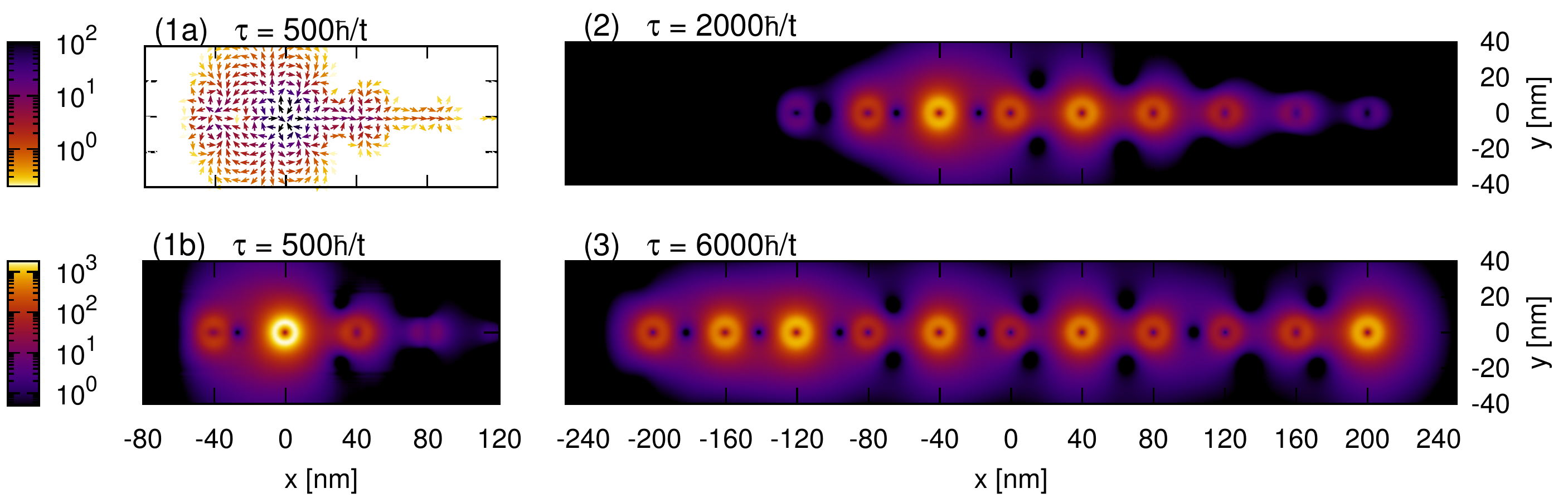} \\
 \caption{(Colour online) Propagation of an electronic excitation along a linear chain of dots. The eleven dots are aligned on a graphene sheet with  $N_z\times N_a=12.000 \times 4.000$ sites corresponding to $(1476 \times 852)$~nm$^2$. The parameters $R$, $D$ and $V$ are chosen as in Fig.~\ref{fig:opt}. At time $\tau=0$ the $a_1$ mode (energy $E=0$) is excited at the central dot. The left panels show the current field (top) and the density profile (bottom) after $\tau=500\hbar/t$. The right panels give two density snapshots at later times.  } 
   \label{fig:linearchain}
\end{figure*}

Finally, we turn to the propagation of an excitation along a linear chain of dots (see Fig.~\ref{fig:linearchain}). For this we consider a configuration of eleven dots in an unbiased graphene sheet. For the applied bias a free-standing dot supports a bound $a_1$ mode. As the wave functions of bound $a_1$ modes at neighbouring dots overlap electrons are no longer permanently confined at one particular dot. Instead, the hybridisation of neighbouring bound states allows the transfer of electrons from one dot to another. We now turn to the propagation of an $a_1$ mode excited at the central dot along the chain of dots. Thereby we discuss the different time scales of electron transport on the lattice scale and between dots.

In Fig.~\ref{fig:linearchain} we plot a consecutive series of three density snapshots. At $\tau=0$ the $a_1$ mode is excited at the central dot. The state we prepare is a superposition of the two states obtained by ED for a single dot which has the current pattern shown in Fig.~\ref{fig:mie_scattering_new}. For comparison with the propagation on the lattice scale we measure time in units of $\hbar/t$. The first snapshot is taken for $\tau=500 \hbar /t$ (on the order of 500 particle hops between neighbouring lattice sites, corresponding to about 106.5 nm of a propagating wave packet). The major fraction of the wave function is still localised at the central dot which shows the density and current density profile for the $a_1$ mode. Obviously the electron transfer between dots is much slower than the transport on the lattice scale. The electron transfer to neighbouring dots proceeds along the preferred scattering directions of the $a_1$ mode. Forward scattering populates the dot to the right while scattering to the left and right backward direction---bent to the neighbouring dot---transfers electron density to the dot on the left.

The difference in inter-dot hopping to the left and right is even more drastic after  $\tau=2000 \hbar /t$. The direct transfer to the right allows a much faster propagation to the right compared with the propagation to the left which is much slower. Note also, that the density at the central dot has decreased and is surpassed by the density at the dots to the right and left.

\section{Conclusions} 

Employing exact numerical techniques for a tight-binding Hamiltonian we have studied the electron dynamics in graphene with circular gate-defined quantum dots. Tracing the time evolution of wave packet scattering on a free-standing dot we find temporary particle trapping at the dot when normal modes become resonant. Thereby we confirm results obtained with the Dirac equation. Signatures of dot-bound states also appear in the optical conductivity. Varying the chemical potential---controlled by the back gate---should allow an experimental confirmation of the series of quasi-bound states at the dot. Following the propagation of an excitation along a linear chain of dots we can identify an effective inter-dot hopping on a reduced time scale.
\vspace*{-0.4cm}

\acknowledgments
This work was supported by the DFG Priority Programme
1459 Graphene. 

\end{document}